\documentclass[a4paper]{jpconf}

\usepackage{hyperref,bm,amsmath,amssymb,a4wide,graphicx,subfigure,wasysym,amsfonts,cite}

\begin{document}

\title{Reconnection of magnetic fields in neutron stars driven by the electron mass term in the triangle anomaly}

\author{Maxim Dvornikov and Victor B. Semikoz}

\address{Pushkov Institute of Terrestrial Magnetism, Ionosphere and Radiowave Propagation (IZMIRAN), 108840 Troitsk, Moscow, Russia}


\ead{maxdvo@izmiran.ru}

\begin{abstract}
The Adler-Bell-Jackiw anomaly for massive particles is studied in an external magnetic field. The contributions of the mean spin and the pseudoscalar are accounted for in the quasiclassical approximation. The equation for the evolution of the magnetic helicity with the new quantum corrections is derived. We show that the quantum contribution to the helicity evolution can overcome the classical one in the dense degenerate matter which can be present in the core of a neutron star. The application of the obtained results for the interpretation of magnetar bursts are discussed.  
\end{abstract}

In the wake of the formulation of the chiral magnetohydrodynamics (MHD)~\cite{SonSur09}, its applications to the studies of various effects in astrophysics~\cite{Mas18}, cosmology~\cite{Sch18} and accelerator physics~\cite{Hir17} are developed. In the majority of situations, the masses of fermions are neglected in the studies of chiral phenomena. Recently, the effect of a nonzero mass on the generation of an electric current along the magnetic field, i.e. the chiral magnetic effect (the CME), was analyzed in our works~\cite{Dvo16PLB,Dvo18}.

In the present work, we summarize our recent results~\cite{DvoSem18} on the fermion mass contribution to the Adler-Bell-Jackiw anomaly, which is known to be closely related to the CME~\cite{NieNin83}. We start with the Adler anomaly for massive fermions and calculate all the terms in the presence of an external magnetic field. Then, we derive the evolution equation for the magnetic helicity, which accounts for the new quantum corrections. We are particularly interested in the ultrarelativistic degenerate plasma, which can be encountered in the core of a neutron star (NS). In the situation, when the chiral imbalance is washed out, e.g., because of interparticle collisions, we compare the classical and quantum surface terms in the magnetic helicity evolution. The quantum contribution is revealed to be dominant in some cases. Finally, astrophysical applications of the obtained results for the explanation of magnetar bursts~\cite{MerPonMel15} are briefly discussed.

We start with the Adler-Bell-Jackiw anomaly for massive fermions in the general form~\cite{Iof06},
\begin{equation}\label{anomgen}
  \partial_\mu
  \left(
    \bar{\psi} \gamma^\mu \gamma^5 \psi
  \right) =
  2\mathrm{i} m \bar{\psi} \gamma^5 \psi +
  \frac{e^2}{8\pi^2} F_{\mu\nu} \tilde{F}^{\mu\nu},
\end{equation}
where $\psi$ is the bispinor of an electron with the nonzero mass $m$, $\gamma^\mu = (\gamma^0,\bm{\gamma})$ and $\gamma^5 = \mathrm{i} \gamma^0 \gamma^1 \gamma^2 \gamma^3$ are the Dirac matrices, $F_{\mu\nu} = (\mathbf{E},\mathbf{B})$ is the electromagnetic field tensor, $\tilde{F}_{\mu\nu} = -\tfrac{1}{2}\varepsilon_{\mu\nu\alpha\beta}F^{\alpha\beta}=(-\mathbf{B},\mathbf{E})$ is the dual tensor, and $e>0$ is the absolute value of the elementary charge.

Decomposing the bispinor into the right and left components $\psi = \psi_\mathrm{R} + \psi_\mathrm{L}$, we rewrite equation~\eqref{anomgen} in the form,
\begin{equation}\label{anomcom}
  \partial_t 
  \left(
    \psi^\dag_\mathrm{R} \psi_\mathrm{R} - \psi^\dag_\mathrm{L} \psi_\mathrm{L}
  \right) +
  \nabla
  \left(
  \psi^\dag \bm{\Sigma} \psi
  \right) =
  2\mathrm{i} m \bar{\psi} \gamma^5 \psi +
  \frac{e^2}{2\pi^2} (\mathbf{E}\cdot\mathbf{B}),
\end{equation}
where $\bm{\Sigma} = \gamma^0\bm{\gamma}\gamma^5$ are the Dirac matrices. After averaging equation~\eqref{anomcom} in electron gas, we get that $\langle\psi^\dag_\mathrm{R} \psi_\mathrm{R}\rangle - \langle\psi^\dag_\mathrm{L} \psi_\mathrm{L}\rangle = n_\mathrm{R} - n_\mathrm{L} \equiv n_5$ is the chiral imbalance, where $n_\mathrm{R,L}$ are the densities of right and left electrons, and $\langle \psi^\dag \bm{\Sigma} \psi \rangle = \bm{\mathcal{S}}$ is the mean spin of electrons.

The mean spin of the electron gas having the chemical potential $\chi$ and the temperature $T$ was obtained in~\cite{SemVal97} as
\begin{equation}\label{meanspin}
  \bm{\mathcal{S}} = 
  - \frac{e\mathbf{B}}{2\pi^2}
  \int_0^{\infty}\mathrm{d}p
  \left\{
    \frac{1}{\exp[(\sqrt{p^2 + m^2} -\chi)/T] + 1}-
    \frac{1}{\exp[(\sqrt{p^2 + m^2} +\chi)/T] + 1}
  \right\}.
\end{equation}
In the case of ultrarelativistic electron gas with $\max(\chi,T) \gg m$ the mean spin in equation~\eqref{meanspin} takes the form,
\begin{equation}\label{chiral-limit}
  \bm{\mathcal{S}} = -
  \frac{e\chi{\bf B}}{2\pi^2}.
\end{equation}
The result in equation~\eqref{chiral-limit} was also known as the chiral separation effect~\cite{MetZhi05} since the mean spin coincides with the spatial components of the axial current, $\psi^\dag \bm{\Sigma} \psi = \bar{\psi} \bm{\gamma} \gamma^5 \psi \equiv \mathbf{J}_\mathrm{A}$.

The calculation of the averaged pseudoscalar $\langle \bar{\psi} \gamma^5 \psi \rangle$ in the external magnetic field was performed in our work~\cite{DvoSem18}. Assuming that the magnetic field is relatively weak, $eB \ll \mathcal{E}_p^2$, where $\mathcal{E}_p = \sqrt{m^2 + p^2}$ is the electron energy, and using the Wentzel-Kramers-Brillouin approximation, we obtain the following result:
\begin{equation}\label{gamma5S5}
  2\mathrm{i}m_e\int \frac{\mathrm{d}^3x}{V}
  \langle \bar{\psi}\gamma_5\psi\rangle =
  - \oint_S \frac{\mathrm{d}^2S}{V} (\bm{\mathcal{S}}_5 \cdot {\bf n}),
\end{equation}
where ${\bf n}$ is unit vector of the external normal to the surface $S$, $V$ is the normalization volume, and
\begin{equation}\label{S5def}
  \bm{\mathcal{S}}_5=-\int\frac{\mathrm{d}^3 p}{\gamma(2\pi)^3}
  \left[
    {\bf S}_e ({\bf p},{\bf x},t) - {\bf S}_{\bar{e}}({\bf p},{\bf x},t)
  \right]
  \left(
    \frac{2}{3} + \frac{1}{3\gamma}
  \right),
\end{equation}
is the new pseudovector. In equation~\eqref{S5def}, $\gamma = \mathcal{E}_p/m$ is the Lorentz factor and ${\bf S}_{e,\bar{e}}({\bf p},{\bf x},t)$ are the spin distribution functions of electrons and positrons.

Note that the calculation of the pseudoscalar $\bar{\psi}\gamma^5\psi$ in parallel electric and magnetic fields was made in~\cite{CopFukPu18}. The relation of the chirality generation with the pairs production was studied in~\cite{CopFukPu18}.

The spin distribution function of the electron gas in the state of equilibrium in the external magnetic field has the form~\cite{DvoSem18},
\begin{equation}\label{spindistr}
  \mathbf{S}_\mathrm{e}^{(\mathrm{eq})}=
  \frac{\mu_\mathrm{B}\mathbf{B}}{\gamma}
  \frac{\mathrm{d}f_\mathrm{e}^{(\mathrm{eq})}}{\mathrm{d}\mathcal{E}_p},
\end{equation}
where $f_\mathrm{e}^{(\mathrm{eq})} = \{\exp[\beta(\mathcal{E}_p-\chi)]+1\}^{-1}$
is the equilibrium Fermi-Dirac distribution, $\beta = 1/T$ is the reciprocal temperature, and $\mu_\mathrm{B} = e/2m$ is the Bohr magneton. We again use the weak magnetic field approximation to derive equation~\eqref{spindistr}. The equilibrium spin distribution function for positrons $\mathbf{S}_{\bar{e}}^{(\mathrm{eq})}$ can be obtained analogously.

The scalar product $(\mathbf{E}\cdot\mathbf{B})$ in equation~\eqref{anomcom} can be rewritten in terms of the magnetic helicity variation. The magnetic helicity $H$ is defined as
\begin{equation}\label{maghel}
  H = \int \mathrm{d}^3 x (\mathbf{A}\cdot\mathbf{B}),
\end{equation}
where $\mathbf{A}$ is the vector potential of the electromagnetic field: $\mathbf{B} = (\nabla \times \mathbf{A})$. Basing on equation~\eqref{maghel} and accounting for the Maxwell equation, $\partial_t \mathbf{B} = - (\nabla \times \mathbf{E})$, one obtains that the magnetic helicity density, $h=H/V$, evolves as~\cite{Priest}
\begin{equation}\label{helevol}
  \frac{{\rm d}h}{{\rm d}t} =
  - 2\int \frac{\mathrm{d}^3x}{V}({\bf E}\cdot{\bf B}) +
  \oint_S \frac{\mathrm{d}^2S}{V} ({\bf n}\cdot 
  \left[
    A_0 {\bf B} +({\bf E}\times {\bf A})
  \right]),
\end{equation}
where $A_0$ is the scalar potential: $\mathbf{E} = - \partial_t \mathbf{A} - \nabla A_0$.

Taking into account equations~\eqref{anomcom}, \eqref{gamma5S5}, and~\eqref{helevol}, one gets the modified equation for the magnetic helicity evolution, which accounts for the quantum contributions~\cite{DvoSem18},
\begin{equation}\label{new_law}
 \frac{{\rm d}}{{\rm d}t}
  \left(
    n_\mathrm{R} - n_\mathrm{L} + \frac{\alpha_\mathrm{em}}{\pi}h
  \right) =
  - \frac{\alpha_\mathrm{em}}{\pi}
  \oint_S \frac{\mathrm{d}^2S}{V} ({\bf n}\cdot 
  \left[
    A_0 {\bf B} +({\bf E}\times {\bf A})
  \right]) - 
  \oint_S
  \frac{\mathrm{d}^2S}{V}
  ({\bf n}\cdot [ \bm{\mathcal{S}} + \bm{\mathcal{S}}_{5}]),
\end{equation}
where $\alpha_\mathrm{em} = e^2/4\pi \approx 1/137$ is the fine structure constant.

The effective mean spin of the electron gas $\bm{\mathcal{S}}_\mathrm{eff} = \bm{\mathcal{S}} + \bm{\mathcal{S}}_{5}$ can be derived on the basis of equations~\eqref{meanspin}, \eqref{S5def}, and~\eqref{spindistr} as
\begin{align}\label{quantum_correction}
  \bm{\mathcal{S}}_\mathrm{eff} = &
  - \frac{e{\bf B}}{2\pi^2}\int_0^{\infty}\mathrm{d}p
  \left[
    1 -
    \frac{1}{3\gamma^2}
    \left(
      \frac{2}{\gamma} + \frac{2}{\gamma^2} - 1
    \right)
  \right]
  \notag
  \\
  & \times
  \left\{
    \frac{1}{\exp [\beta(\mathcal{E}_p - \chi)] +1} -
    \frac{1}{\exp [\beta(\mathcal{E}_p + \chi)] +1}
  \right\}.
\end{align}
Using equation~\eqref{quantum_correction}, one gets that the combined spin effect practically vanishes in a nonrelativistic plasma, $\bm{\mathcal{S}}_\mathrm{eff}=\bm{\mathcal{S}} + \bm{\mathcal{S}}_5  \to 0$ since $\mathcal{E}_p=\gamma m \approx m$. In the case of the ultrarelativistic electron gas with $\gamma\gg 1$, one obtains from equation~(\ref{quantum_correction}) that $\bm{\mathcal{S}}_\mathrm{eff} \approx \bm{\mathcal{S}}= - e\chi {\bf B}/2\pi^2$; cf. equation~(\ref{chiral-limit}). Hence the magnetization effect is great under the condition $\chi \gg m$, which can be implemented, e.g., in the core of NS.

To have a nonzero contribution of the quantum surface terms to the magnetic helicity evolution in equation~\eqref{new_law}, one should suppose that the chemical potential $\chi$ is coordinate dependent. Otherwise, $\oint_S \mathrm{d}^2 S ({\bf n}\cdot \bm{\mathcal{S}}_\mathrm{eff}) \equiv 0$ because of the Gauss theorem. The dependence $\chi(\mathbf{x})$ can be accounted for if we recall that the number density of the degenerate electron gas in an external magnetic field has the form~\cite{Nunokawa:1997dp},
\begin{equation}\label{totaldensity}
  n_e=\frac{e B p_{\mathrm{F}e}}{2\pi^2} + \sum_{n=1}^{n_\mathrm{max}}
  \frac{2eB\sqrt{p_{\mathrm{F}e}^2 - 2eBn}}{2\pi^2},
\end{equation}
where $p_{\mathrm{F}e}$ is the Fermi momentum of the electron gas and $n_\mathrm{max}$ is the maximal Landau number, which depends on the integer part of $p_{\mathrm{F}e}^2/(2eB)$.

Let us discuss a moderately strong  magnetic field with
$m^2\ll 2eB\ll p_{\mathrm{F}e}^2$, that is valid for charged particles within NS obeying $p_{\mathrm{F}e}=p_{\mathrm{F}p}\sim 100\,{\rm MeV}$, where $p_{\mathrm{F}p}$ is the Fermi momentum of protons. Here we assume that a plasma is electrically neutral, $n_e=n_p$. The above condition is fulfilled for magnetars having $B\sim 10^{15}\,{\rm G}$~\cite{MerPonMel15}. Using equation~\eqref{totaldensity}, one obtains the correction to the number density of electrons at the lowest Landau level linear in the magnetic field strength~\cite{Nunokawa:1997dp},
\begin{equation}\label{real}
  n_e\approx \frac{p_{\mathrm{F}e}^3}{3\pi^2}
  \left[
    1 + \frac{3eB}{2p_{\mathrm{F}e}^2}
  \right].
\end{equation}
If we consider the correction to the electron number density from the magnetic field in equation~\eqref{real}, $\chi$ in equation~\eqref{quantum_correction} becomes dependent on the magnetic field. Hence the chemical potential is coordinate dependent in case of a nonuniform magnetic field.

Let us consider the evolution of the magnetic field in the dense degenerate matter of NS. The electron gas is ultrarelativistic in this case, as mentioned above. The chiral imbalance in such a system is washed out very rapidly because of electron-proton collisions. The typical relaxation time for $n_5$ was found in~\cite{Dvo16,Dvo17} to be $10^{-11}\,\text{s}$ for $T\sim 10^8\,\text{K}$. Thus we should set $n_\mathrm{R} = n_\mathrm{L}$ in equation~\eqref{new_law}. The magnetic field in NS should be taken as axially symmetric. Hence, using equation~\eqref{real} one gets that
\begin{equation}\label{anisotrop}
  \chi(r,\theta)=[3\pi^2n_e(r)]^{1/3}
  \left\{
    1 - \frac{eB(r,\theta)}{2[3\pi^2n_e(r)]^{2/3}}
  \right\},
\end{equation}
where $n_e(r) = n_\mathrm{core}Y_e(1 - r^2/R^2_\mathrm{NS})$~\cite{Lattimer}, $n_\mathrm{core} \sim 10^{38}\,\text{cm}^{-3}$ is the density in the NS center, $R_\mathrm{NS}\sim 10\,\text{km}$ is the typical NS radius, and $Y_e\sim(\text{several percent})$ is the electron fraction. In equation~\eqref{anisotrop}, we use the spherical coordinates $r$ and $\theta$.

Then, we use equations~\eqref{new_law} and~\eqref{anisotrop}, as well as take into account the Ohm law ${\bf E}= - {\bf v}\times{\bf B} + {\bf j}/\sigma_\mathrm{cond}$, where ${\bf j}$ is the electric current and $\sigma_\mathrm{cond}$ is the electric conductivity. Assuming that $\sigma_\mathrm{cond}$ is great in the NS matter, we get that only the surface terms contribute to the magnetic helicity evolution~\cite{DvoSem18},
\begin{align}
  \frac{\mathrm{d}H}{\mathrm{d}t}= &
  \left(
    \frac{\mathrm{d}H}{\mathrm{d}t}
  \right)_\mathrm{class} +
  \left(
    \frac{\mathrm{d}H}{\mathrm{d}t}
  \right)_\mathrm{quant},
  \label{dHdt}
  \\
  \left(
    \frac{\mathrm{d}H}{\mathrm{d}t}
  \right)_\mathrm{class} = &
  \oint_S\mathrm{d}^2S
  [({\bf v}\cdot{\bf n})({\bf A}\cdot{\bf B}) -
  ({\bf v}\cdot{\bf A})({\bf n}\cdot{\bf B})],
  \label{dHdtclass}
  \\
  \left(
    \frac{\mathrm{d}H}{\mathrm{d}t}
  \right)_\mathrm{quant} = &
  \frac{1}{\chi(R)}
  \oint_S\mathrm{d}^2S B(R,\theta)({\bf n}\cdot {\bf B}).
  \label{dHdtquant}
\end{align}
Equation~\eqref{dHdtclass} is well known in classical MHD~\cite{Priest}. The new quantum contribution in equation~\eqref{dHdtquant} is given by the chemical potential $\chi(R)=[3\pi^2n_e(R)]^{1/3}$ at the distance $R$ from the NS center.

\begin{figure}
  \centering
  \includegraphics[scale=0.4]{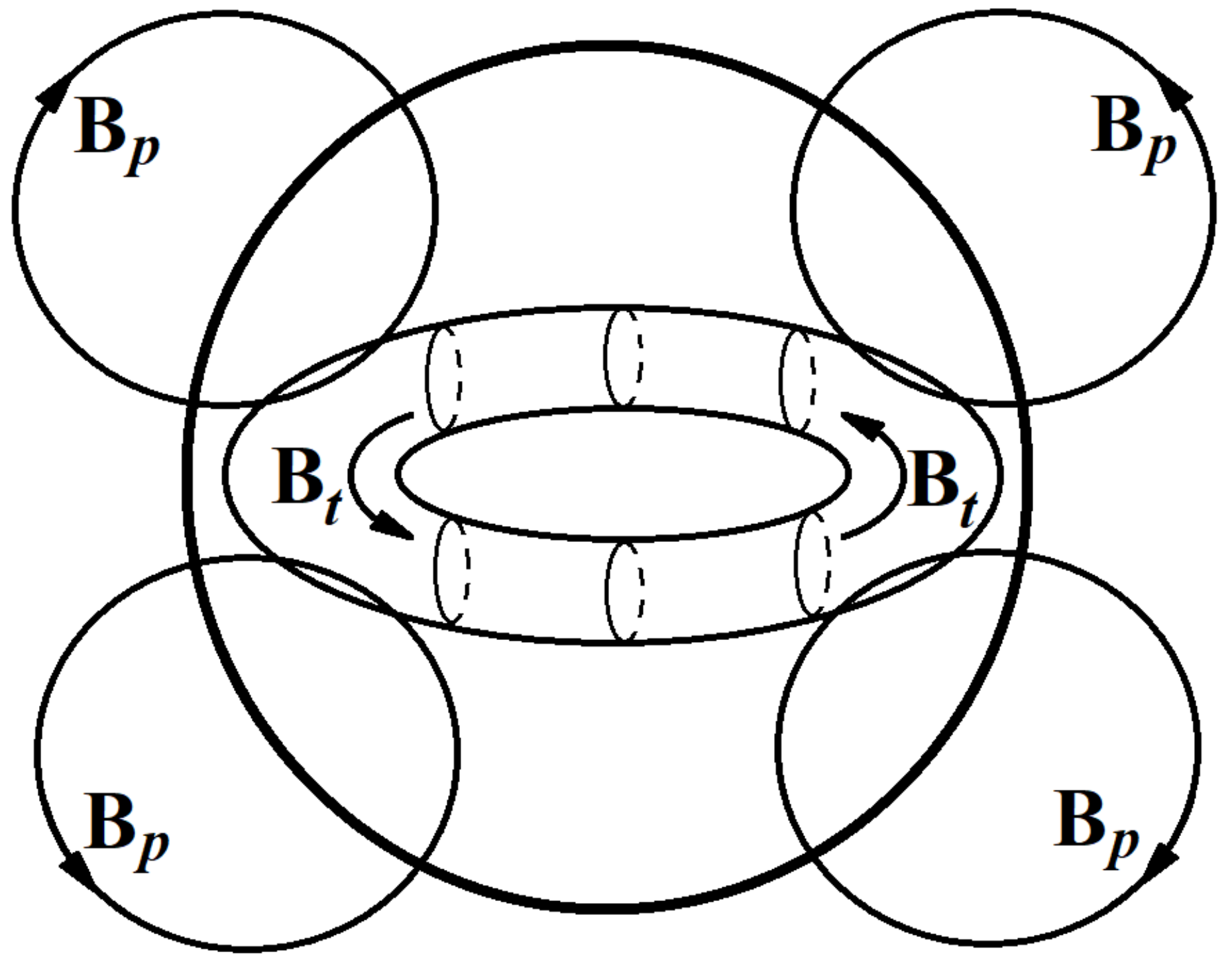}
  \caption{The schematic illustration of the magnetic field configuration given
  in equation~\eqref{model}. The toroidal component is
  $\mathbf{B}_t = B_{\varphi}(r)\cos \theta {\bf e}_{\varphi}$ and the poloidal field reads
  $\mathbf{B}_p = B_{p}(r)[\cos 2\theta {\bf e}_r + \sin 2\theta {\bf e}_{\theta}]$.  
  \label{fig:magfields}}
\end{figure}

Let us choose the quadrupole configuration of the magnetic field as
\begin{equation}\label{model}
  {\bf B}(r,\theta)= B_{p}(r)[\cos 2\theta {\bf e}_r + \sin 2\theta {\bf e}_{\theta}] +
  B_{\varphi}(r)\cos \theta {\bf e}_{\varphi},
\end{equation}
where $B_{p}$ and $B_{\varphi}$ are the amplitudes of the poloidal and the toroidal components, and the angle $\theta$ is measured from the equator of NS. The structure of the magnetic field in equation~\eqref{model} is schematically illustrated in figure~\ref{fig:magfields}.

For the magnetic field given in equation~\eqref{model} the quantum contribution to the magnetic helicity evolution in equation~\eqref{dHdt} can be estimated as~\cite{DvoSem18}
\begin{equation}\label{quantum}
  \left(
    \frac{{\rm d}H}{{\rm d}t}
  \right)_\mathrm{quant} = 
  \frac{2\pi}{\chi} R^2B_p B_{\varphi}
  A
  \sim
  \frac{B^2R^2}{\chi},
\end{equation}
where $A = A(B_p,B_{\varphi})$ is a given function of the poloidal and toroidal magnetic field components.

We found in~\cite{DvoSem18} that the quantum contribution in equation~\eqref{quantum} to the magnetic helicity evolution can be greater than the classical one in equation~\eqref{dHdtclass} in the dense matter in NS if we consider almost rigid rotation of a star in a corotating frame. The variation of the magnetic helicity leads to the change of the magnetic field topology. This change can be interpreted as the reconnection of magnetic field lines. The phenomenon of the magnetic reconnection is known to result in the dissipation of the magnetic energy~\cite{Priest}. This transformation of the magnetic energy can have the implication to electromagnetic bursts of magnetars~\cite{MerPonMel15}.

\ack

One of the authors (M.D.) is thankful to 
RFBR (research project No.~18-02-00149a) for a partial support.

\end{document}